\documentclass{article}

\usepackage{spconf,amsmath,graphicx}

\usepackage[hidelinks]{hyperref}

\usepackage{url}
\usepackage{tabularx}
\usepackage{multirow}
\usepackage{float}
\usepackage{caption}
\usepackage{booktabs}
\usepackage{subcaption}

\title{Comparison of Self-Supervised Speech Pre-Training Methods on Flemish Dutch}
\name{Jakob Poncelet, Hugo Van hamme}
\address{KU Leuven \\ 
     Department Electrical Engineering ESAT-PSI\\
     Kasteelpark Arenberg 10, Bus 2441, B-3001 Leuven Belgium\\
     \textit{\{jakob.poncelet, hugo.vanhamme\}@esat.kuleuven.be}}

\begin{document}

\maketitle

\begin{abstract}
Recent research in speech processing exhibits a growing interest in unsupervised and self-supervised representation learning from unlabelled data to alleviate the need for large amounts of annotated data. We investigate several popular pre-training methods and apply them to Flemish Dutch. We compare off-the-shelf English pre-trained models to models trained on an increasing amount of Flemish data. 
We find that the most important factors for positive transfer to downstream speech recognition tasks include a substantial amount of data and a matching pre-training domain. Ideally, we also finetune on an annotated subset in the target language. All pre-trained models improve linear phone separability in Flemish, but not all methods improve Automatic Speech Recognition. We experience superior performance with wav2vec 2.0 and we obtain a 30\% WER improvement by finetuning the multilingually pre-trained XLSR-53 model on Flemish Dutch, after integration into an HMM-DNN acoustic model. 
\end{abstract}


\section{Introduction}
Building a good speech recogniser typically requires a large amount of annotated data from a specific language. Obtaining high-quality labelled data is a costly and time-intensive process, and for many languages this remains a big issue. However, even in a highly resourced language like English, recent work has shown impressive results in Automatic Speech Recognition (ASR) by pre-training on unlabelled data and transferring that knowledge to regular speech recognition models \cite{baevski2020wav2vec,xu2020selftraining,zhang2020pushing} 
or even completely unsupervised speech recognition \cite{baevski2021unsupervised}. 
This paradigm shift towards unlabelled data is of great significance as untranscribed recordings of speech are much easier to acquire. 

Self-supervised learning is a clever way to learn general information from data without requiring any labels. Recently many successful methods have emerged for self-supervised representation learning from speech. The general idea is to implicitly learn the global structure and local characteristics that are inherently present in speech. Depending on the task, both local information, such as the pronunciation of a specific phoneme, and more global information, such as speaker traits and recording properties, can be useful. By pre-training a network with a well-chosen objective function, these relevant attributes about the input speech can be captured and summarised in rich feature vectors. This improves several downstream tasks like speech recognition and typically reduces the amount of required data, since the principal characteristics are already extracted and more easily accessible. Moreover, the structure of a speech waveform is to some extent general and language-independent, which explains the improvements with these features in low-resource languages \cite{conneau2020unsupervised, riviere2020unsupervised}.

The objective function used in self-supervised learning techniques is the driving force behind the extraction of powerful speech representations. 
In fact, the self-supervised objective has more impact on the learned representation than architectural differences between methods \cite{icassp2021_chung}.
In Autoregressive Predictive Coding (APC) \cite{chung2019unsupervised, chung2020generative, chung2020vectorquantized, chung2020improved}, the objective is to predict a frame a few steps ahead, given the information up to that point. Another branch of research focuses on predicting the current frame given past and future context, by reconstructing several masked frames \cite{Liu_2020, chi2021audio}, similar to the Masked Language Modeling (MLM) approach in Natural Language Processing \cite{devlin2019bert}. Finally, Contrastive Predictive Coding (CPC) \cite{oord2019representation} is a popular technique in representation learning, where the objective is to predict the future in the latent space and a contrastive loss is applied to maximise mutual information. CPC has been successfully applied to speech recognition \cite{baevski2020wav2vec, schneider2019wav2vec, baevski2020vqwav2vec, baevski2020effectiveness} and has shown to be able to learn robust and cross-lingual speech representations \cite{conneau2020unsupervised, riviere2020unsupervised, kawakami2020learning}. We refer to the literature for other related work in self-supervised and unsupervised representation learning \cite{Chorowski_2019, khurana2020convolutional, liu2020tera, liu2020nonautoregressive, ravanelli2020multitask, pascual2019learning, jiang2020speech, Ling_2020}.

Following the widespread improvements in ASR as a result of self-supervised pre-training, this paper will focus on Flemish Dutch, a medium-resourced language. Flemish is the language spoken in Flanders, the Dutch-speaking part of Belgium. It is closely related to the Dutch variant spoken in The Netherlands, but there are still many noticeable differences \cite{Velde2010WillDB}. A few seconds of speech suffice to distinguish the two variants. Although geographically relatively small, Flemish Dutch is diverse with several dialects, roughly corresponding to the five provinces in Flanders, though natives will observe even finer detail. Furthermore, Dutch belongs to the family of West-Germanic languages, like English and German, which makes it a very interesting language to examine whether pre-training on English leads to strong improvements in Dutch. While there is some overlap in phones, there are also several vowels and diphthongs that do not occur in English. 

In this work, we compare several popular self-supervised pre-training methods when applied to Flemish. First of all, we look at the applicability of off-the-shelf models that are pre-trained on English and assess the transferability to Flemish. This would be convenient for several research domains and technological applications. Additionally it would eliminate the need for large computational resources necessary to pre-train these models, which scales with the model size (e.g. the high-capacity wav2vec 2.0 model \cite{baevski2020wav2vec}) and for many models also with the amount of data. Second, we examine the importance of matching the pre-training language to the target language as opposed to the amount of data used in pre-training. To this end, we compare pre-trained models in English and Netherlands Dutch to models trained on Flemish Dutch. Recent work \cite{conneau2020unsupervised} has shown that low-resource languages can greatly benefit from higher-resource languages when they are more similar due to positive transfer, but cross-lingual representation learning degrades the performance on high-resource languages due to interference. 
Furthermore, self-supervised pre-training has shown to improve robustness and reduce the degradation on out-of-domain data \cite{robustW2V2, ma2021probing}. We show that simply augmenting the finetuning data leads to a strong speech recognition improvement in noisy and reverberated environments.
Finally, we investigate ASR improvements with the recent wav2vec 2.0 model \cite{baevski2020wav2vec, conneau2020unsupervised} and study several pre-training and finetuning scenario's with an increasing amount of data, yielding substantial reduction of Word Error Rates (WER) compared to the baselines. 

We evaluate the models in terms of linear phone separability by reporting the classification accuracy of an external linear classifier. The classifier is trained to predict Flemish Dutch phones from the features extracted from each model. For the ASR experiments, we report the results of an HMM-DNN hybrid model \cite{Povey_ASRU2011} where the DNN is trained with the learned features.

\section{Models}
The procedure consists of three separate phases: 1) pre-training a model on data without labels, 2) optionally finetuning the model on a labelled set with transcripts, 3) extracting the learned features to perform a downstream evaluation task.

\subsection{Self-Supervised Pre-training}
We start with a short description of the investigated pre-training techniques and refer to the corresponding papers for more details. Table~\ref{tab:overview} gives an overview of all models.

\begin{table*}[htbp]
  \caption{Shallow overview and comparison of all pre-training techniques.}
  \label{tab:overview}
  \centering
  \small
  \begin{tabularx}{\textwidth}{| l || X | X | X | X | p{2.3cm} |}
    \toprule
    \textbf{Model} & \textbf{Feature encoder} & \textbf{Aggregator} & \textbf{Objective} & \textbf{Output dimension} & \textbf{\# Parameters}\\
    \midrule
    \textbf{APC} & Filterbank & GRU & Reconstruct future frame & 512 & 4.1M \\
    \hline
    \textbf{Mockingjay} & Filterbank & Bidirectional Transformer & Reconstruct masked frame & 768 & 21.3M \\
    \hline
    \textbf{CPC} & CNN & LSTM & Identify future feature & 256 & 1.8M \\
    \hline
    \textbf{wav2vec} & CNN & CNN & Identify future feature & 512 & 32.5M \\
    \hline
    \textbf{wav2vec 2.0} & CNN & Transformer & Identify quantised future feature & 768 (base), 1024 (large)  & 95.0M (base), 317.3M (large) \\
    \bottomrule 
  \end{tabularx}
\end{table*}

\subsubsection{APC}
In APC, autoregressive models encode the temporal information in the past sequence of frames, for example with Gated Recurrent Units (GRU). A future frame, $n$ steps ahead of the current frame, is linearly predicted from the autoregressive outputs. The model is then trained with an L1 reconstruction loss on the predicted frame. We use a model with 3 GRU layers and predict 5 steps ahead \cite{chung2019unsupervised, chung2020vectorquantized}. The outputs of the last GRU layer are extracted as features for the downstream task.

\subsubsection{Mockingjay}
While APC conditions its prediction on past context only, Mockingjay leverages both past and future context to predict a frame that has been masked out. The encoder is a deep bidirectional Transformer \cite{vaswani2017attention} that learns contextualised representations, which are extracted from the last layer. These representations are linearly mapped to predict the masked frames, and the model is trained with a reconstruction loss between the predicted and true frames. We use the base model with 3 Transformer blocks in the encoder \cite{Liu_2020, S3PRL}.

\subsubsection{CPC}
CPC directly applies a stack of strided convolutional layers to the raw waveform to encode the sequence in the latent space. An autoregressive model (the aggregator) then looks at the representations of the past sequence and its output is mapped to predict the latent representations for several steps in the future. The loss is not reconstructive, but contrastive: given the aggregator output, the model has to distinguish the correct sample out of a bunch with distractors from windows more distant in time or from different sequences. We use the modified CPC approach \cite{riviere2020unsupervised} where the encoder exists of 5 CNN layers, the autoregressive model is an LSTM and the prediction network is a 1-layer Transformer network. The model predicts 1 to 12 steps in the future, with a separate projection layer for every step, and is trained with 10 distractors. The outputs of the autoregressive model are the extracted features. 

\subsubsection{wav2vec}
Wav2vec is built on CPC but uses a fully convolutional model. The autoregressive model is replaced by a context network consisting of 12 convolutional layers. Two additional linear transformations increase the capacity of the encoder (this architecture is called \textit{wav2vec large} in the corresponding paper \cite{schneider2019wav2vec}). The outputs of the context network are the feature vectors \cite{ott2019fairseq}.

\subsubsection{wav2vec 2.0}
Wav2vec 2.0 combines ideas from wav2vec \cite{schneider2019wav2vec}, vq-wav2vec \cite{baevski2020vqwav2vec} and MLM. The encoder computes latent speech representations from the raw waveform with 7 temporal convolution blocks. A certain proportion of the latent features is masked before feeding to the aggregator, which is a Transformer network. At the same time, a quantisation module maps the latent feature vectors to discretised versions. The final training objective is then to distinguish the true quantised representation for a masked time step, given the aggregator output \cite{baevski2020wav2vec}. We differentiate between the base and large architecture of the model, which contain respectively 12 and 24 Transformer blocks in the aggregator. The contextual features at the output of the aggregator are extracted for downstream tasks \cite{ott2019fairseq}. We duplicate them in time to mimic a stride of 10ms instead of 20ms. 

The wav2vec 2.0 model can be finetuned on a labelled set. To this end, an extra linear layer is added on top of the context network and a CTC loss is applied with the transcription characters as targets. The encoder is frozen during finetuning. Finetuning is done after the pre-training is completed. 

Finally, XLSR-53 is a large wav2vec 2.0 model pre-trained on 53 languages simultaneously \cite{conneau2020unsupervised}. The authors have shown that the quantised speech representations can express connections between languages when trained in a multilingual setup.

Due to limited resources, we pre-train wav2vec 2.0 base models for 100k updates and finetune for 500k updates, and we don't pre-train our own wav2vec 2.0 large models but only finetune existing pre-trained models.

\subsection{Downstream Feature Evaluation}
\subsubsection{Phone Classification}
\label{sec:phoneclass}
We train an external phone classifier consisting of just one linear layer and a softmax layer \cite{S3PRL}, with as input the features extracted from the pre-trained models. All pre-trained features are compared to the baseline of 80-dimensional log-mel filterbank features, including second order delta features and mean-variance normalisation. For every utterance, there is a phone label every 10ms, corresponding to the stride of the input features. The classifier is trained with a cross-entropy loss. We report the accuracy of the classifier of predicting the correct phone label for every 10ms window, instead of using the most voted phone during its entire duration, because the learned representations should contain phonetic information even at the start of a phone. 

For English experiments we use the phone labels from \cite{oord2019representation}, which have been generated by forced alignment with Kaldi \cite{Povey_ASRU2011} using pre-trained models on LibriSpeech, and mapped to 41 classes. For Flemish experiments we use the phone labels provided in the Corpus Gesproken Nederlands (Section~\ref{sec:CGN}). The phone sequences have been computed by forced alignment on the manually checked orthographic transcripts with SPRAAK \cite{demuynck_laureys}, and have been partly manually checked as well. There are 49 distinct phone classes \cite{CGN_Oostdijk}.

\subsubsection{ASR}
\label{sec:asrmet}
We train a baseline HMM-DNN model with Kaldi \cite{Povey_ASRU2011} on MFCC features. The HMM-GMM models triphones and includes LDA, MLLT and fMLLR transformations. It is trained on MFCC features to compute alignments and build a phonetic tree with one state per phone. For the pre-trained models, we reuse the alignments and tree from the MFCC model and only train the DNN model with the extracted features as input. We make a distinction between a large DNN model containing 14 TDNN-F layers \cite{Povey2018} (similar to the Switchboard recipe) and a small DNN model with only 3 TDNN-F layers. We leave out iVector extraction and speed perturbation, and remove the delta layers for pre-trained features. We decode with a pruned trigram language model and use a lexicon of 100k words. We report Word Error Rates based on the Levenshtein distance, but make a correction for inconsistencies in compounding (which occur frequently in Dutch).

\section{Data}
\subsection{Flemish Dutch datasets}
\subsubsection{Labelled data}
\label{sec:CGN}
Corpus Gesproken Nederlands (CGN) \cite{CGN_Oostdijk} - also called Spoken Dutch Corpus - is a manually annotated speech database of around 900 hours of Dutch, of which 270 hours correspond to Flemish Dutch. CGN contains both phonetical and word-level transcriptions and segmentations. The labelled data can be used for finetuning, for ASR model training and for the proposed evaluation procedures. We make the distinction between three training sets of data, based on the type of speech. 

\textbf{VL-train-clean} This set contains 35h of prepared, read speech by professional readers. This corresponds to component O of CGN. 

\textbf{VL-train-other} This set contains several types of speech, including read speech (\textit{VL-train-clean}), news reports, interviews, lectures, sports commentary, etc. This set holds 145h of data from components B,F,G,H,I,J,K,L,M,N,O of CGN.

\textbf{VL-train-all} This set contains all components from the CGN database and corresponds to 270 hours of speech. The difference with \textit{VL-train-other} is the inclusion of narrowband telephone speech (8kHz resampled to 16kHz) and spontaneous conversational speech, which correspond to respectively components C,D and component A of CGN.

In a similar way, we make a distinction between \textbf{VL-test-clean} (4h) and \textbf{VL-test-other} (15h, including the 4h from \textit{VL-test-clean}). There is no overlap in speakers with the train sets.

For phone classification experiments in English, we use the \textit{train-clean-100} set of LibriSpeech \cite{librispeech} and use the train-test split and phone labels from \cite{oord2019representation}.

\subsubsection{Unlabelled data}
We have created a dataset of 450h of unlabelled data for unsupervised experiments in Flemish Dutch, by extracting audio from online available resources. We refer to this set as \textbf{VL-unsup}. This set consists of 200h of data from recordings in the Flemish parliament, 100h of audio from broadcast TV news and 150h of audio from TV talkshows. For pre-training, we use this set and the labelled sets without the transcriptions.

\subsection{Pre-trained models}
\label{sec:prm}
For some experiments, we use off-the-shelf available pre-trained models for APC, Mockingjay, CPC, wav2vec and wav2vec 2.0 \cite{S3PRL, ott2019fairseq}. These models have been pre-trained on English audiobooks from LibriSpeech (\textit{LS-960}) \cite{librispeech}, LibriLight (\textit{LL-60k}) \cite{librilight} or both, i.e. LibriVox (\textit{LV-60k}) \cite{Pratap_2020}. The XLSR-53 model is trained on 56k hours of data from 53 different languages. The XLSR data originates from CommonVoice \cite{ardila2020common}, Multilingual LibriSpeech \cite{Pratap_2020},  and BABEL \cite{babel}. It includes around 1.6k hours of Dutch \cite{conneau2020unsupervised} of which we recon only a very small part is Flemish Dutch (a few hours in CommonVoice). We also use a wav2vec 2.0 model pre-trained on the Dutch part of VoxPopuli (\textit{VP-NL-4.5k}) \cite{voxpopuli}, which contains 4.5k hours of Netherlands Dutch speech recordings from the European parliament.

\section{Discussion}
\subsection{Phone classification}
\subsubsection{Applicability of off-the-shelf models to Flemish Dutch}
First, we perform phone classification as explained in Section~\ref{sec:phoneclass}. For experiments in English, we train and test the classifier on a train-test split of LibriSpeech \textit{train-clean-100}. For experiments in Flemish, we either train a classifier on \textit{VL-train-clean} and test on \textit{VL-test-clean}, or train on \textit{VL-train-other} and evaluate on \textit{VL-test-other}. Table~\ref{tab:prep} shows the phone classification accuracies with features extracted from English pre-trained models that are online available (see Section~\ref{sec:prm}).

\begin{table}[hbt]
    \centering
    \caption{Linear phone classification accuracy (\%) with features extracted from off-the-shelf models pre-trained on English. We evaluate classification on English and Flemish.}
    \footnotesize
    \begin{tabular}{l|c|c|c}
    \toprule
    \multirow{2}{*}{\textbf{Model}} & \textbf{English} & \multicolumn{2}{c}{\textbf{Flemish}} \\
    & \textit{LS-tc100} & \textit{VL-test-clean} & \textit{VL-test-other} \\
    \midrule
    Baseline & 48.0 & 48.5 & 39.3 \\
    APC & 72.7 & 71.4 & 60.1 \\
    Mockingjay & 68.1 & 71.4 & 59.1 \\
    CPC & 71.3 & 71.7 & 60.5 \\
    wav2vec & \textbf{78.4} & \textbf{73.3} & \textbf{62.4} \\
    wav2vec 2.0 (base) & 75.1 & 71.7 & 58.8 \\
    \bottomrule
    \end{tabular}
    \label{tab:prep}
\end{table}

\noindent The relative improvements with respect to the baseline as a result of pre-training are consistent across both languages. The accuracy on \textit{VL-test-clean} is of a similar magnitude as the accuracy on \textit{LS-tc100}, which can be explained by the fact that both sets contain rather easy, clean speech. On \textit{VL-test-clean} and \textit{VL-test-other}, we see absolute accuracy improvements of more than 20\%. This shows that the pre-training techniques improve linear phone separability, even when the target language differs from the pre-training language.  

\subsubsection{Language Matching}
Second, we examine the effect of matching the domain (i.e. the language, but also the type of speech) of the pre-training speech to the target speech. We pre-train models on Flemish Dutch data, compare them to other pre-trained models, and investigate the effect of finetuning wav2vec 2.0 on a Flemish subset. Table~\ref{tab:pca} reports phone classification accuracies for pre-training and finetuning on several datasets. 

\begin{table}[ht]
    \centering
    \caption{Phone classification accuracy (PCA) percentage when training a classifier on \textit{VL-train-clean} and testing on \textit{VL-test-clean} ('clean'), and when training a classifier on \textit{VL-train-other} and testing on \textit{VL-test-other} ('other').}
    \footnotesize
    \begin{tabularx}{\columnwidth}{X|X|X|p{0.5cm}|p{0.5cm}}
        \toprule
        \multirow{2}{\hsize}{\textbf{Model}} & \multirow{2}{\hsize}{\textbf{Pre-training}} &
        \multirow{2}{\hsize}{\textbf{Finetuning}} & \multicolumn{2}{c}{\textbf{PCA}} \\
        & & & clean & other \\
        \midrule
        Baseline (Filterbank) &  -- & -- & 48.5 & 39.3 \\
        \hline
        \multirow{5}{\hsize}{APC} & LS-960 & -- & 71.4 & 60.1 \\
        \cline{2-5}
        & VL-train-clean & -- & 66.9 & 54.8 \\
        \cline{2-5}
        & VL-train-other & -- & 67.6 & 57.1 \\
        \cline{2-5}
        & VL-unsup & -- & \textbf{73.3} & \textbf{63.3} \\
        \cline{2-5}
        & VL-train-all + VL-unsup & -- & \textbf{73.3} & 63.0 \\
        \hline
        \multirow{2}{\hsize}{Mockingjay} & LS-960 & -- & 71.4 & 59.1 \\
        \cline{2-5}
        & VL-train-clean & -- & 65.2 & 53.5 \\
        \hline
        \multirow{4}{\hsize}{CPC} & LL-60k & -- & 71.7 & \textbf{60.5} \\
        \cline{2-5}
        & VL-train-clean & -- & \textbf{72.6} & 55.6 \\
        \cline{2-5}
        & VL-train-other & -- & 69.3 & 59.5 \\
        \cline{2-5}
        & VL-unsup & -- & 66.8 & 57.4 \\
        \cline{2-5}
        & VL-train-all + VL-unsup & -- & 67.5 & 58.1 \\
        \hline
        wav2vec & LS-960 & -- & 73.3 & 62.4 \\
        \hline
        \multirow{5}{\hsize}{wav2vec 2.0 base} & LS-960 & -- & 71.7 & 58.8 \\
        \cline{2-5}
        & VP-NL-4.5k & -- & 64.5 & 50.0 \\
        \cline{2-5}
        & VL-train-other & -- & 47.4 & 36.1 \\
        \cline{2-5}
        & VL-train-all + VL-unsup & -- & 54.7 & 43.9 \\
        \cline{2-5}
        & LS-960 & VL-train-other & \textbf{83.6} & \textbf{76.2} \\
        \cline{2-5}
        & VP-NL-4.5k & VL-train-other & \textbf{83.6} & 76.1 \\
        \cline{2-5}
        & VL-train-other & VL-train-other & 81.3 & 74.1 \\
        \cline{2-5}
        & VL-train-all + VL-unsup & VL-train-other & 82.2 & 75.0 \\
        \hline
        \multirow{7}{\hsize}{wav2vec 2.0 large} & LS-960 & -- & 55.9 & 45.2 \\
        \cline{2-5}
        & LV-60k & -- & 24.6 & 14.3 \\
        \cline{2-5}
        & XLSR-53 & -- & 34.4 & 21.8 \\
        \cline{2-5}
        & VP-NL-4.5k & -- & 58.2 & 45.7 \\
        \cline{2-5}
        & LS-960 & VL-train-other & 81.2 & 73.4 \\
        \cline{2-5}
        & LV-60k & VL-train-other & 85.0 & 76.6 \\
        \cline{2-5}
        & XLSR-53 & VL-train-other & \textbf{86.4} & \textbf{79.1} \\
        \cline{2-5}
        & VP-NL-4.5k & VL-train-other & 84.12 & 76.3 \\
        \bottomrule
    \end{tabularx}
    \label{tab:pca}
\end{table}

For APC, we notice an improvement over the English pre-trained model when we match the training and target language and use a sufficient amount of data (but still less than LibriSpeech). For CPC, we experienced converging difficulties and a high sensitivity to the number of training cases. We see an improvement over the pre-trained model on \textit{VL-test-clean} when only training on \textit{VL-train-clean}. This might suggest that domain matching is important for CPC. The LibriLight pre-trained model is trained on much more data (60k hours), which can explain the strong performance on \textit{VL-test-other}. 

For wav2vec 2.0, the base models trained on Flemish data and the pre-trained model on VoxPopuli perform worse than the LibriSpeech model, despite matching language (VL) or using more data in a related language (VP). The former is most likely explained by sub-optimal training, the latter could be explained by the fact that the VoxPopuli parliament recordings reflect different acoustic conditions to certain components with clean speech in CGN. Finetuning on Flemish leads to very high phone classification accuracies for all models. 

\noindent For the large high-capacity wav2vec 2.0 models, we note low accuracies without finetuning. 
Other works corroborate this finding and ascribe it to the problem-agnostic pre-training, and have shown that certain audio features are more easily accessible from the middle layers in very deep transformer models than the output layer \cite{baevski2021unsupervised, ma2021probing}. Finetuning (with graphemic transcripts) alleviates this discrepancy and gives an improved accuracy over the base models with the large models. It seems that the large model also benefits more from a large amount of pre-training data, as the LibriVox and XLSR model show.

The XLSR-53 model reports the highest score. It is trained on a similar amount of data as the LibriVox model, but the training data contains Dutch, German and English. 
We also note that the (Flemish) Dutch stops, which are prevoiced voiced stops, differ from the aspirated voiceless stops in English. This is also better covered in the XLSR-53 set.

\subsection{ASR Results}
\subsubsection{Clean ASR}
Table~\ref{tab:asr} reports WER results on \textit{VL-test-other} of HMM-DNN ASR experiments with a large DNN and features from different models, as described in Section~\ref{sec:asrmet}. Every ASR model (including the baseline model) is trained on \textit{VL-train-other}.

\begin{table}[ht]
    \centering
    \caption{ASR experiments with large DNN ASR model, reporting WER on \textit{VL-test-other}.}
    \footnotesize
    \begin{tabularx}{\columnwidth}{X|X|X|p{0.65cm}}
        \toprule
        \textbf{Model} & \textbf{Pre-training} & \textbf{Finetuning} & \textbf{WER} \\
        \midrule
        Baseline (MFCC) & -- & -- & 15.10 \\
        \hline
        \multirow{3}{*}{APC} & LS-960 & -- & 16.02 \\
        & \multirow{2}{\hsize}{VL-train-all + VL-unsup} & \multirow{2}{*}{--} & \multirow{2}{*}{16.20} \\
        & & & \\
        \hline
        CPC & LS-960 & -- & 15.03 \\
        \hline
        wav2vec & LS-960 & -- & 14.89 \\
        \hline
        \multirow{3}{\hsize}{wav2vec 2.0 base} & LS-960 & -- & 13.84 \\
        \cline{2-4}
        & VL-train-other & -- & 14.44 \\
        \cline{2-4}
        & VL-train-all + VL-unsup & -- & 13.52 \\
        \cline{2-4}
        & LS-960 & VL-train-other & \textbf{11.42} \\
        \cline{2-4}
        & VL-train-other & VL-train-other & 13.41 \\
        \cline{2-4}
        & VL-train-all + VL-unsup & VL-train-other & 11.76 \\
        \hline
        \multirow{6}{\hsize}{wav2vec 2.0 large} & LS-960 & -- & 14.33 \\
        \cline{2-4}
        & LV-60k & -- & 14.72 \\
        \cline{2-4}
        & VP-NL-4.5k & -- & 16.32 \\
        \cline{2-4}
        & XLSR-53 & -- & 13.40 \\
        \cline{2-4}
        & LS-960 & VL-train-other & 10.87 \\
        \cline{2-4}
        & LV-60k & VL-train-other & 12.65 \\
        \cline{2-4}
        & XLSR-53 & VL-train-other & \textbf{10.61} \\
        \bottomrule
    \end{tabularx}
    \label{tab:asr}
\end{table}

\noindent The improvements compared to the baseline with MFCC are small, if any, except for wav2vec 2.0. In contrast to the phone classification experiments, we see an improvement over the LibriSpeech model (960h) when we pre-train the base model on a comparable amount of Flemish (720h) without finetuning, which supports the idea that the combination of a matching pre-training language and a large amount of data is key.

\noindent The poor performance of the large VoxPopuli model can possibly be explained by a poor transfer from Netherlands Dutch to Flemish and different acoustic conditions compared to the test set. Also, contrarily to phone classification, the large wav2vec 2.0 model pre-trained on LibriSpeech outperforms the model pre-trained on LibriVox.

Analogous to the phone classification experiments, the XLSR-53 model - which has a large variability in its training data - yields a significant WER improvement, manifesting a positive cross-lingual transfer to Flemish, and the best results are obtained when finetuning a model on an annotated Flemish subset. We obtain almost 30\% relative WER improvement when finetuning XLSR-53 on Flemish, compared to the baseline. The difference with the LibriSpeech model is however small. We postulate that a large wav2vec 2.0 model trained on more Flemish data would equal or improve this result. 

\subsubsection{Effect of amount of pre-training and finetuning data}
We quantitatively examine the effect of an increasing amount of data used for pre-training wav2vec 2.0 base models (unlabelled) or finetuning XLSR-53 (labelled). We shuffle all available data from all sets. We also evaluate finetuning on different types of speech (from different sets). For the unlabelled dataset, this distinction is not trivial. Table~\ref{tab:asr2} shows the results. The ASR model is always trained on \textit{VL-train-other}.

\begin{table}[ht]
    \caption{WER with DNN ASR model in function of the amount of Flemish data used for pre-training or finetuning.}
    \label{tab:asr2}
    \begin{subtable}{\columnwidth}
    \centering
    \footnotesize
    {\begin{tabularx}{\columnwidth}{X|X|X|X|X|X|X|X|X}
        \toprule
        10h & 30h & 50h & 100h & 150h & 250h & 350h & 500h & 700h \\
        \hline
        31.87 & 20.85 & 16.76 & 15.55 & 15.74 & 14.76 & 14.34 & 14.73 & 13.52 \\
        \bottomrule
    \end{tabularx}}
    \caption{Unlabelled data for pre-training a base wav2vec 2.0 model (no finetuning), large ASR DNN.}
    \label{pt1}
    \end{subtable}%
    
    \begin{subtable}{\columnwidth}
    \centering
    \footnotesize
    {\begin{tabularx}{\columnwidth}{X|X|X|X|X|X|X|X|X}
        \toprule
        0h & 1h & 10h & 20h & 30h & 50h & 90h & 150h & 250h \\
        \hline
        27.75 & 13.84 & 12.08 & 11.32 & 11.19 & 10.71 & 10.61 & 10.53 & 10.50 \\
        \bottomrule
    \end{tabularx}}
    \caption{Labelled data for finetuning XLSR-53, small ASR DNN.}
    \label{tab:ft1}
    \end{subtable}%
    
    \begin{subtable}{\columnwidth}
    \centering
    \footnotesize
    {\begin{tabularx}{\columnwidth}{p{0.8cm}|p{1.3cm}|p{1.4cm}|X|X}
    \toprule
    VL set & No FT (0h) & Clean (29h) & Other (128h) & All (248h) \\
    \hline
    WER & 13.40 & 12.35 & 10.61 & 10.58 \\ 
    \bottomrule
    \end{tabularx}}
    \caption{Different sets of CGN for finetuning XLSR-53, large ASR DNN.}
    \label{tab:ft2}
    \end{subtable}
\end{table}

\noindent For pre-training, it is necessary to have a considerable amount of data to improve upon the baseline, and more Flemish data gives improvements. It seems that the learned audio representations include acoustic details aside from more abstract phoneme qualities, as the WER on 150h of shuffled data is higher than when pre-training on an equal amount of matched data (\textit{VL-train-other} in Table~\ref{tab:asr}). This might suggest a high dependency on acoustic conditions. For finetuning, the data should match the type or conditions of the test set for optimal results, and the improvements saturate with more data. Note that a small DNN suffices after finetuning, but a large DNN is required when the XLSR model is not finetuned (first column of Table~\ref{tab:ft1}). This is in line with the poor phone classification results of the large models without finetuning.

\subsubsection{ASR in noisy environments}

We investigate the robustness of wav2vec 2.0 to noisy and reverberated speech by replicating the \textit{VL-test-other} set in 4 different scenario's: filtered with RIRs (\textit{rev}), with added noise at a certain SNR (\textit{noise1}: 5-20dB, \textit{noise2}: 0-15dB) and both (\textit{rev} + \textit{noise3}: 5-15dB). We use noises from multiple sources (NTT Noise-DB, CHIME2, NoiseX, DEMAND, Humming) and RIRs from the Aachen Impulse Response Database. We compare wav2vec 2.0 large models pre-trained on LibriVox: without finetuning, finetuned on \textit{VL-train-other} ('clean') and finetuned on a fourfold augmented \textit{VL-train-other} ('aug') by adding noise and reverberation in Table~\ref{tab:noisy}.

\begin{table}[ht]
    \centering
    \caption{WER with large DNN ASR on augmented \textit{VL-test-other} with LibriVox pre-trained large wav2vec 2.0 models.}
    \footnotesize
    \begin{tabularx}{\columnwidth}{X|X|X|X|X|X|X}
        \toprule
        \multirow{2}{\hsize}{\textbf{Model}} & \multirow{2}{\hsize}{\textbf{FT}} & \multicolumn{5}{c}{\textbf{WER}} \\
        & & \textit{clean} & \textit{rev} & \textit{noise1} & \textit{noise2} & \textit{rev + noise3} \\
        \midrule
        MFCC & -- & 15.10 & 28.36 & 20.58 & 26.26 & 39.21 \\
        \hline
        w2v2 & -- & 14.71 & 27.12 & 19.96 & 25.28 & 39.19 \\
        \hline
        w2v2 & clean & 12.43 & 22.60 & 16.31 & 20.61 & 33.32 \\
        \hline
        w2v2 & aug & 12.13 & 18.08 & 14.64 & 17.39 & 24.43 \\
        \bottomrule
    \end{tabularx}
    \label{tab:noisy}
\end{table}

\noindent Finetuning on augmented data gives strong improvements over finetuning on clean data in the reverberated and noisy settings, with 3-9\% absolute WER reduction. More so, there is even a slight improvement in the clean setting as well, probably because of more finetuning data.

\section{Conclusion}
Pre-trained features on English speech transfer well to Flemish Dutch in terms of improving linear phone separability. These self-supervised pre-trained models are readily available and easy to use. Matching the pre-training and target language further improves results, but either matching the type of speech or using a larger amount of data is necessary. The recently proposed wav2vec 2.0 model appears superior, especially when finetuned on data from the target language. Finally, we obtain the best results with the large multilingually trained XLSR-53 model and see nearly 30\% improvement in WER by finetuning the XLSR-53 model on Flemish, compared to the baseline. We show the importance of matching pre-training and target language and acoustic conditions.

\section{Acknowledgements}
This research received funding from the Flemish Government under the "Onderzoeksprogramma Artificiële Intelligentie (AI) Vlaanderen" programme.

\pagebreak

\bibliographystyle{IEEEbib}

\bibliography{main}

\end{document}